# Techniques for modeling a high-quality *B*-spline curves by *S*-polygons in a float format

Rushan Ziatdinov[1], Valerijan Muftejev[2], Rifkat Nabiyev[2], Albert Mardanov[2], Rustam Akhmetshin[2]

ziatdinov@kmu.ac.kr|muftejev@mail.ru|dizain55@yandex.ru|systems@list.ru|ahmetshin@bk.ru

[1]Department of Industrial and Management Engineering, Keimyung University, Daegu, Republic of Korea;

[2]Department of the Fundamentals of Mechanisms and Machines Design, Ufa State Aviation Technical University, Ufa, Russian Federation

*This article proposes a technique for the geometrically stable modeling of high-degree B-spline curves based on S-polygon in a float format, which will allow the accurate positioning of the end points of curves and the direction of the tangent vectors. The method of shape approximation is described with the purpose of providing geometrical proximity between the original and approximating curve. The content of the notion of a "harmonious, regular form" of B-spline curve's S-polygon in a float format is revealed as a factor in achieving a high-quality of fit for the generated curve. The expediency of the shape modeling method based on S-polygon in a float format at the end portions of the curve for quality control of curve modeling and editing is substantiated. The results of a comparative test are presented, demonstrating the superlative efficacy of using the Mineur-Farin configuration for constructing constant and monotone curvature curves based on an S-polygon in a float format. The findings presented in this article confirm that it is preferable to employ the principle of "constructing a control polygon of a harmonious form (or the Mineur-Farin configuration) of a parametric polynomial" to a B-spline curve's S-polygon in a float format, and not to a B-polygon of the Bézier curve. Recommendations are given for prospective studies in the field of applying the technique of constructing a high-quality B-spline curves to the approximation of log-aesthetic curves, Ziatdinov's superspirals, etc. The authors of the article developed a technique for constructing smooth connections of B-spline curves with ensuring a high order of smoothness of the composite curve. The proposed techniques are implemented in the FairCurveModeler program as a plug-in to engineering CAD systems.*

**Keywords:** *S-polygon, NURBS, fair curve, typical curve, transition curve.*

> *"The Golden Age will include people who will learn to unite."*
> **Konstantin Tsiolkovsky**

## 1. Introduction

Mineur et al. [1] and Farin [2] proposed a method for constructing Bézier curves with a monotone change in curvature, based on the construction of the Bézier polygon (also known as *B*-polygon) with a monotonic change in the control polygon legs at fixed angles between the legs. The error in this method was recently demonstrated [18], although this fact was known a decade ago to one of our Japanese colleagues, Norimasa Yoshida[1].

In this paper, the aforementioned method adapts to the construction of *B*-spline curves with monotone curvature and the application of the *S-polygon* with the *Mineur-Farin configuration*. A technique for modeling a curve with an *S*-polygon in a float format is proposed, which allows for the precise positioning of the end points of the curve and the direction of the tangents. More detailed equivalence parameters for the two curves, such as the similarity of the *m*-level curvature and the similarity of the *m*-level curvature of torsion, are introduced. In addition, a method for approximating analytic curves with the equivalence of curvature and curvature of torsion to the *m*th level is proposed.

Modern CAD systems use a method of modeling *class A curves* using *S*-polygons of high-degree NURBS curves. To model closed curves, the *S*-polygon is used in a float format, and the closed format of the *S*-polygon is used for non-closed curves. Based on the configuration of the *S*-polygon in a float format, the quality of the modeled curve is easily judged. The harmonious, regular form of the *S*-polygon in a float format of the *B*-spline curve allows us to ensure high quality in the generated curve (hereinafter in this article we will clarify the concept of harmony of the *S*-polygon shape). Using a closed *S*-polygon is dictated by the need for precise positioning of the end points of the curve and the directions of the tangent vectors at the end points. The end portions of the closed *S*-polygon coincide with the initial portions of the *B*-polygons of the end segments of the curve, which allows precise positioning of the end points and directions of the tangent vectors at these points. However, using an *S*-polygon in a closed format causes instability in the curve modeling at the end portions of the curve.

## 2. Techniques for modeling B-spline curves

The Cox-de Boor algorithm [4, 5] using the format of the representation of the *B*-spline curve with an *S*-polygon in a float format is numerically stable, since it uses a sequence of operations dividing the segment into a given ratio. In addition, the operations of calculating points and differential characteristics of the curve are uniform on any segment. Obviously, for the best approximation of the circular arc, the segment of the *B*-spline curve of the vertex of the *S*-polygon in a float format must be uniformly chosen on the circle. We construct an *S*-polygon in a float format of a segment of a *B*-spline of the ninth degree along the vertices of a regular dodecagon inscribed in a circle with a radius of $R = 10$. The point of the largest deviation of the evolute of the curve from the center $(0, 0)$ of the circle is the point with the coordinates $(0.000000015321, 0.000000057206)$.

Transformations of the control polygon formats of *B*-spline curves and Bézier splines are performed using BS and SB algorithms [6-8]. In [9] NBS and NSB algorithms of transformations for rational splines are proposed. SB (NSB) algorithms for the transition from the format of the *S*-polygon *B*-spline to the format of the generalized *B*-polygon of the Bézier spline are stable, since they contain only operations of dividing the segments into a given ratio. BS (NBS) algorithms are less stable, so they contain extrapolation operations of a segment in a given ratio. The algorithms for converting *S*-polygon formats of multi-segment splines partially contain the operations of BS (NBS) and SB (NSB) algorithms.

---

[1] Private communication.

The publication of the conference proceedings was supported by the RFBR, grant No. 18-07-20045\18

Fig. 1 shows the segment of the *B*-spline curve of the ninth degree associated with the *S*-polygon in a float format and the closed *S*-polygon coinciding with the *B*-polygon of the Bézier curve of the ninth degree.

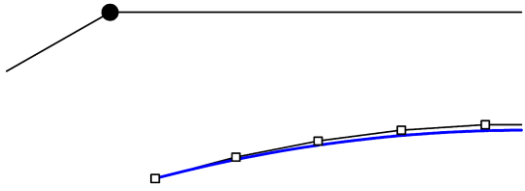

**Fig. 1.** An *S*-polygon in a float format and a closed *S*-polygon of a segment of a *B*-spline curve of ninth degree.

The vertices of the closed *S*-polygon (i.e., the *B*-polygon) do not lie on the circle. The legs of the *S*-polygon do not have the same length (the lengths of the legs are 0.518838, 0.519949, 0.520746, 0.521227, 0.521387, ...). The angles between the legs are also not equal: in radians they are 0.0653566, 0.0654361, 0.0654898, 0.0655169, ...). At the same time, it is this configuration that provides a high-precision approximation of the circular arc with an accuracy of $5.92221 \times 10^{-8}$.

In modern CAD systems for modeling class A curves, *B*-splines with *S*-polygons in closed format are used. It is impossible to draw a closed *S*-polygon with the same configuration as in Fig. 2 to approximate a circular arc with the same level of accuracy. The difficulty of ensuring proper quality at the end portions is emphasized in the guidelines for modeling class A curves in CAD systems.

## 3. Technique of geometrically stable modeling of high-degree B-spline curves using an S-polygon in a float format

An *S*-polygon in a float format of a *B*-spline curve of degree *m* can be used for quality control of modeling and for editing a curve [10].

Differential characteristics of the spline are determined by the divided differences of the *S*-polygon in a float format [5, 11]. A harmonious, regular form of an *S*-polygon in a float format of a *B*-spline curve of degree *m* must satisfy the second order isogeometry and regularity of an *S*-polygon of order *m*-1 [9]. Such a configuration of an *S*-polygon can be achieved by providing the required law of variation of the discrete approximation of the tangent vectors and curvature [9], as computed from the central divided differences of the *S*-polygon of the first and second order. More stringent requirements to the harmonicity of the polyline configuration can be introduced if one takes into account deviations that portions of a modeled or approximated curve with a monotone change in curvature are determined by segments of a polygonal line with a monotonous variation of the discrete curvature, as, for example, segments of a polygonal line with a Mineur-Farin configuration.

Modeling and editing of the high-degree *B*-spline curve at the end portions of the curve by means of a closed *S*-polygon is extremely unstable. Even simple rounding of the coordinates of a closed *S*-polygon's vertices can cause distortion of the shape. For example, let the arc of a *B*-spline curve of the ninth degree be given by a convex *S*-polygon in a float format (see Figure 1), built on the vertices of a dodecagon inscribed in a circle of radius *R* = 10. Rounding the coordinates of the second vertex of the closed *S*-polygon (see Fig. 1) from the value (-1.807034341228000, 8.748581800695999) to the value (-1.807, 8.748) causes a significant change in the shape of the curve and an unpredictable change in the shape of the *S*-polygon in a float format. The vertices of the original *S*-polygon (-7.071, -7.071), (-9.659, -2.588), (-9.659, 2.588), etc. have changed to (-83.972,

1295.780), (-8.274, -26.046), (-9.832, 5.520), and so on. The error of circular arc approximation increases from $5.92221 \times 10^{-8}$ to a value of $3.06354 \times 10^{-1}$.

It is the stability of shaping with an *S*-polygon in a float format at the end portions of the curve that makes this method preferable.

The uncertainty of positioning the end points of the curve and the values of the tangent vectors is eliminated by employing a technique which consists of the following steps:

1. Construct the *S*-polygon of a curve in a float format of the proper shape.
2. Control the positions of the end points and the tangent directions when formatting the closed *S*-polygon.
3. Define the position difference vector (the vector determined by the difference of the actual end point versus the desired end point) and the difference angle of the direction (the angle between the actual tangent direction and the required one).
4. Switch the current formatting to that of *S*-polygon in a float format.
5. Eliminate misalignment in the position of the end points, requiring translation of the *m* terminal vertices of the *S*-polygon by the magnitude of the difference vector in the direction of decreasing the error value.
6. Eliminate the difference in the direction of the tangent vectors, requiring rotation of *m* end vertices of the *S*-polygon by the amount of the difference angle in the direction of decreasing the difference value.
7. Edit the changed configuration of the *S*-polygon in a float format until it reaches a harmonious, regular shape. Repeat the position control on the closed *S*-polygon and edit the *S*-polygon in a float format until a harmonious, regular form of the *S*-polygon in a float format and the accuracy of the closed *S*-polygon position are obtained.

## 4. Typical curves

In the works of Mineur et al. [1] and Farin [2] algorithms for generation of the control points for the formation of *typical Bézier curves* with a monotone change in curvature were developed. In these algorithms, a *B*-polygon of the Bézier curve is generated as the control polygon. The essence of the algorithms lies in the construction of a *B*-polygon with a monotonous change in the length of the leg, while maintaining a constant value of the angle between the legs and the angle of torsion. Such a configuration of a *B*-polygon is called the *Mineur-Farin configuration*.

The principle of constructing a control polygon by a Mineur-Farin configuration is correct in itself, however, as will be shown below, it is more appropriate to apply it to an *S*-polygon in a float format of the *B*-spline curve, and not to the *B*-polygon of the Bézier curve.

We will perform a comparative testing of these two different approaches for the limiting case, namely, modeling the curve of constant curvature. Obviously, the elongation coefficient of the leg in the Mineur-Farin configuration must be equal to 1 for generating the curve of constant curvature. It is also obvious that the Bézier curve and polynomial splines do not accurately approximate the geometry of the circular arc. However, a comparison of the approximation accuracy can demonstrate the advantage of using one method over another.

To generate the vertices of an *S*-polygon with a *B*-spline curve, we use an equilateral polygon inscribed in a circle of radius *R* = 10. For a degree of *n* = 3, a hexagon is used, and for degrees of *n* = {5, 7, 9} a dodecagon is used.

The construction of the vertices of a *B*-polygon is performed as follows: The terminal vertices of the *B*-polygon which coincide with the terminal vertices of the *B*-spline curve segment. The tangent vectors at the endpoints are orthogonal to the radius vectors from the center (0, 0). The intermediate vertices of the *B*-polygon are defined in such a way as to obtain a configuration with a constant leg length and a fixed angle between the legs.

Below is a table with the distance values of the evolute points farthest from the zero center (0, 0), of the Bézier curve and the *B*-spline curve.

Table 1. Comparison of approximation accuracy

| Degree ($n$) | Bézier curve | *B*-spline curve |
|---|---|---|
| 3 | $8.17283 \times 10^{-1}$ | $8.33333 \times 10^{-1}$ |
| 5 | $8.06272 \times 10^{-2}$ | $1.11371 \times 10^{-3}$ |
| 7 | $4.65855 \times 10^{-2}$ | $7.85284 \times 10^{-6}$ |
| 9 | $3.19524 \times 10^{-2}$ | $5.92221 \times 10^{-8}$ |

As can be seen from Table 1, an increase in the degree results in an extremely slow increase in the approximation accuracy of a circular arc by Bézier curve. Even for a degree of $n = 9$, the Bézier curve does not provide acceptable accuracy (in modern CAD systems, the NURBS degree does not exceed a tenth degree). Concurrently, for $n = 9$, the *B*-spline curve achieves accuracy acceptable for engineering calculations in CAD systems. This test result shows the advantage of using the Mineur-Farin method for constructing a curve of constant curvature by means of an *S*-polygon in a float format. This statement is also true in the case of modeling curves with a monotone change in curvature. Consider the configuration of a polygon of five points with a monotonous change in the length of the leg with a scaling factor of two units and a fixed angle between the legs equal to 90°. Let us define the Bézier curve of the fifth degree on this configuration and on the same configuration, as on the *S*-polygon in the float format, we define the segment of the *B*-spline curve. Analysis of the evolute graphs shows that the curvature of the Bézier curve is not monotonic, and the *B*-spline curve is defined with a monotone change in the curvature (Fig. 2).

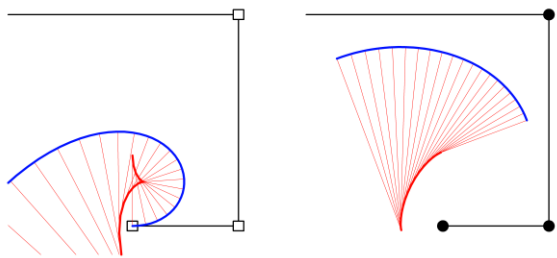

**Fig. 2.** Bézier curve (left) and *B*-spline curve (right) segments, defined on the same control polygon.

## 5. Construction of a high-quality approximating B-spline curve

Welters and Farin [3] proposed an approximation method taking advantage of the geometric proximity of two curves (the original curve and the approximating curve). In Welters and Farin's method, geometric proximity is estimated by examining the relative equivalence of two curves with respect to the parameters of curvature and torsion. Moreover, the authors introduced the term *shape approximation*, and noted that successful approximation guarantees the accuracy of the shape in the sense that the geometry of the original curve is approximated neatly [3].

To approximate the form, we propose to use additional, highly detailed characteristics of the internal geometry of the curve, proposed in [12]:

- Curvature of the first level is the curvature function itself $\kappa(s)$.
- The torsion of the first level is the torsion function itself $\tau(s)$.
- The second level curvature is a function of curvature of the curvature function $\kappa^{(2)}(s) = (\kappa \circ \kappa)(s)$
- The second level curvature of the torsion is a function of curvature of the torsion: $\kappa_\tau^{(2)}(s) = (\kappa \circ \tau)(s)$.

We define of the curvature of the *n*th level from the curvature of a curve as follows
$$\kappa^{(n)}(s) = \left(\kappa^{(n-1)} \circ \kappa\right)(s),$$
and the curvature of the *n*th level from the torsion of the curve is
$$\kappa_\tau^{(n)}(s) = \left(\kappa_\tau^{(n-1)}(s) \circ \tau\right)(s).$$

Under shape preserving approximation of *n*-level of equivalence we mean the equivalence of curvatures to the *n*-th level and of the curvatures of torsion to the *n*-th level for the original curve and the approximating curve. Equivalence of the shape of the curves means that the number and order of inflection and extrema points in the curves are the same. Thus, for high-quality shape approximation, the curvature of the curvature and the curvature of the torsion of the original and approximating curve should be equivalent in shape, at least until the second level. These arguments lead us to the following technique of isogeometric shape approximation of an analytic curve by *B*-spline curve of high-degree *n*, where *n*>5. The vertices of an *S*-polygon in a float format of a *B*-spline curve are generated directly on the analytic curve. With a proper distribution of points and the distribution density on the analytic curve, it is possible to ensure the equivalence of the shape of the analytic curve segment and the *B*-spline curve not lower than the second level.

As an example, a shape approximation of a conical spiral by a *B*-spline curve of the eighth degree is given (Fig. 3). The conical spiral is determined by the parametric equation
$$x(s) = 2 + s \, \sin s$$
$$y(s) = 2 + s \, \cos s$$
$$z(s) = s$$
The vertices of the *S*-polygon are computed on an analytic curve with the initial value of the parameter $s = 0$ and with a step $h = 1$ in the number of twenty points.

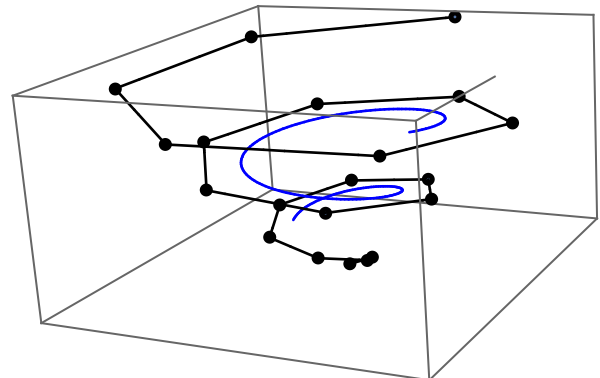

**Fig. 3.** Approximation of the shape of a conical spiral.

The curvature graph as well as the graph of curvature of torsion of the *B*-spline curve up to the second level are equivalent in shape to the second level curvatures of the conical spiral.

This technique will also be useful for approximating the log-aesthetic curves [13, 14] and Ziatdinov's superspirals [15]. In accordance with the multi-criteria approach to the evaluation of

The publication of the conference proceedings was supported by the RFBR, grant No. 18-07-20045\18

log-aesthetic curves [16], when importing log-aesthetic curves to CAD systems in the form of NURBS templates, it is important to preserve the characteristics of the internal geometry using smoothness criteria.

## 6. Composite curves

Existing methods of modern CAD systems provide an order of smoothness of no more than $G^4$. With the float format of the $S$-polygon, it is easy to construct composite log-aesthetic curves by segments of $B$-spline curves of the same degree of $n$ approximating log-aesthetic curves not lower than the second level of shape isogeometry. The $S$-polygons of the $B$-spline curves are positioned and matched by additional vertices so that the composite $S$-polygon has a harmonious, regular shape. Hence, a composite $B$-spline curve will have an order of smoothness of $n$-1.

## 7. Conclusion

This article demonstrates the techniques of applying $S$-polygons in a float format for modeling a high-quality $B$-spline curve using smoothness criteria. In particular, the technique of modeling the non-closed $B$-spline curve is proposed with the provision of accurate positioning of the end points of the curve and the directions of the tangent vectors at the end points. In addition, a technique for modeling a $B$-spline curve with a monotonic curvature through an $S$-polygon with a Mineur-Farin configuration is proposed, as well as a technique for approximating analytic curves with the equivalence of curvature and curvature of torsion to a level $m$. A technique has been developed for constructing smooth connections of $B$-spline curves with a high order of smoothness of the composite curve.

The proposed techniques are implemented in the program FairCurveModeler [17].

## Acknowledgements

This manuscript was copy-edited by volunteers Mr. Duncan Griffiths (Canada), Mr. Anthony Dowds (UK) and Mr. Marty Smith (USA). Additionally, we thank Ms. Kimberley Hoffman (USA), Ms. Linda Sherwood (USA) and Mr. Seth I. Rich (USA) for useful remarks and suggestions. Without their kind assistance, this publication would not be possible.